# Statistical Inverse Problem: Root Approach

Yu. I. Bogdanov

OAO "Angstrem", 124460, Moscow, Russia[1]

**ABSTRACT**

Multiparametric statistical model providing stable reconstruction of parameters by observations is considered. The only general method of this kind is the root model based on the representation of the probability density as a squared absolute value of a certain function, which is referred to as a psi-function in analogy with quantum mechanics. The psi-function is represented by an expansion in terms of an orthonormal set of functions. It is shown that the introduction of the psi-function allows one to represent the Fisher information matrix as well as statistical properties of the estimator of the state vector (state estimator) in simple analytical forms. The chi-square test is considered to test the hypotheses that the estimated vector converges to the state vector of a general population. The method proposed may be applied to its full extent to solve the statistical inverse problem of quantum mechanics (root estimator of quantum states). In order to provide statistical completeness of the analysis, it is necessary to perform measurements in mutually complementing experiments (according to the Bohr terminology). The maximum likelihood technique and likelihood equation are generalized in order to analyze quantum mechanical experiments. It is shown that the requirement for the expansion to be of a root kind can be considered as a quantization condition making it possible to choose systems described by quantum mechanics from all statistical models consistent, on average, with the laws of classical mechanics.

## 1. FISHER INFORMATION MATRIX AND STATE ESTIMATOR

A psi-function considered further is a mathematical object of statistical data analysis. The introduction of the psi-function implies that the "square root" of the probability density is considered instead of the probability density itself.

$$p(x) = |\psi(x)|^2$$

Let the psi function depend on $s$ unknown parameters $c_0, c_1, \ldots, c_{s-1}$ (according to quantum mechanics, the basis functions are traditionally numbered from zero corresponding to the ground state). The parameters introduced are the coefficients of an expansion in terms of a set of basis functions. Assume that the set of the functions is orthonormal.

For the sake of simplicity, consider first a real valued psi function. Let an expansion have the form

$$\psi(x) = \sqrt{1 - (c_1^2 + \ldots + c_{s-1}^2)}\,\varphi_0(x) + c_1 \varphi_1(x) + \ldots + c_{s-1} \varphi_{s-1}(x). \quad (1)$$

Here, we have eliminated the coefficient $c_0 = \sqrt{1 - (c_1^2 + \ldots + c_{s-1}^2)}$ from the set of parameters to be estimated, since it is expressed via the other coefficients by the normalization condition.

The parameters $c_1, c_2, \ldots, c_{s-1}$ are independent. We will study their asymptotic behavior using the Fisher information matrix [1-2]

$$I_{ij}(c) = n \cdot \int \frac{\partial \ln p(x,c)}{\partial c_i} \frac{\partial \ln p(x,c)}{\partial c_j} p(x,c)\,dx$$

---
[1] e-mail: bogdanov@angstrem.ru

The fundamental significance of the Fisher information matrix consists in its property to set the constraint on achievable (in principle) accuracy of statistical estimators. According to the Cramer - Rao inequality [1-2], the matrix $\Sigma(\hat{\theta}) - I^{-1}(\theta)$ is nonnegative for any unbiased estimator $\hat{\theta}$ of an unknown vector valued parameter $\theta$. Here, $\Sigma(\hat{\theta})$ is the covariance matrix for the estimator $\hat{\theta}$. The corresponding difference asymptotically tends to a zero matrix for the maximum likelihood estimators (asymptotic efficiency).

It is of particular importance for our study that the Fisher information matrix drastically simplifies if the psi function is introduced [3-4]

$$I_{ij} = 4n \cdot \int \frac{\partial \psi(x,c)}{\partial c_i} \frac{\partial \psi(x,c)}{\partial c_j} dx. \qquad (2)$$

In the case of the expansion (1), the information matrix $I_{ij}$ is $(s-1) \times (s-1)$ matrix of the form

$$I_{ij} = 4n \left( \delta_{ij} + \frac{c_i c_j}{c_0^2} \right), \quad c_0 = \sqrt{1 - (c_1^2 + \ldots + c_{s-1}^2)}. \qquad (3)$$

A noticeable feature of the expression (3) is its independence on the choice of basis functions. Note that only the representation of the density in the form $p = |\psi|^2$ results in a universal (and simplest) structure of the Fisher information matrix.

In view of the asymptotic efficiency of the maximum likelihood estimators, the covariance matrix of the state estimator is the inverse Fisher information matrix:

$$\Sigma(\hat{c}) = I^{-1}(c) \qquad (4)$$

Let us extend the covariance matrix by appending the covariance between the $c_0$ component of the state vector and the other components. In result, we find that the covariance matrix components are

$$\Sigma_{ij} = \frac{1}{4n} (\delta_{ij} - c_i c_j) \quad i,j = 0,1,\ldots,s-1. \qquad (5)$$

From the geometrical standpoint, the covariance matrix (5) is a second-order tensor. Moreover, the covariance matrix (up to a constant factor) is a single second-order tensor satisfying the normalization condition.

In quantum mechanics, the matrix

$$\rho_{ij} = c_i c_j \qquad (6)$$

is referred to as a density matrix (of a pure state). Thus,

$$\Sigma = \frac{1}{4n} (E - \rho), \qquad (7)$$

where $E$ is the $s \times s$ unit matrix.

In the diagonal representation,

$$\Sigma = U D U^{+}, \qquad (8)$$

where $U$ and $D$ are unitary (orthogonal) and diagonal matrices, respectively.

As is well known from quantum mechanics, the density matrix of a pure state has the only (equal to unity) element in the diagonal representation. Thus, in our case, the diagonal of the $D$ matrix has the only element equal to zero (the corresponding eigenvector is the state vector); whereas the other diagonal elements are equal to $\frac{1}{4n}$ (corresponding eigenvectors and their linear combinations form a subspace that is orthogonal complement to the state vector). The zero element at a principle diagonal indicates that the inverse matrix (namely, the Fisher information matrix of the $s$-th order) does not exist. It is clear since there are only $s-1$ independent parameters in the distribution.

The results on statistical properties of the state vector reconstructed by the maximum likelihood method can be summarized as follows. In contrast to a true state vector, the estimated one involves noise in the form of a random deviation vector located in the space orthogonal to the true state vector. The components of the deviation vector (totally, $s-1$ components) are asymptotically normal independent random variables with the same variance $\frac{1}{4n}$. In the aforementioned $s-1$-dimensional space, the deviation vector has an isotropic distribution, and its squared length is the random variable $\frac{\chi^2_{s-1}}{4n}$, where $\chi^2_{s-1}$ is the random variable with the chi-square distribution of $s-1$ degrees of freedom, i.e.

$$1-\left(c,c^{(0)}\right)^2 = \frac{\chi^2_{s-1}}{4n}. \tag{9}$$

This expression means that the squared scalar product of the true and estimated state vectors is smaller than unity by asymptotically small random variable $\frac{\chi^2_{s-1}}{4n}$.

The results found allow one to introduce a new stochastic characteristic, namely, a confidence cone (instead of a standard confidence interval). Let $\vartheta$ be the angle between an unknown true state vector $c^{(0)}$ and that $c$ found by solving the likelihood equation. Then,

$$\sin^2\vartheta = 1 - \cos^2\vartheta = 1-\left(c,c^{(0)}\right)^2 = \frac{\chi^2_{s-1}}{4n} \leq \frac{\chi^2_{s-1,\alpha}}{4n}. \tag{10}$$

Here, $\chi^2_{s-1,\alpha}$ is the quantile corresponding to the significance level $\alpha$ for the chi-square distribution of $s-1$ degrees of freedom.

The set of directions determined by the inequality (10) constitutes the confidence cone. The axis of a confidence cone is the reconstructed state vector $c$. The confidence cone covers the direction of an unknown state vector at a given confidence level $P=1-\alpha$.

Root estimator provides refined representations of such classical results as chi-squared criterion and Gaussian approximation of binomial distribution.

Let $p_1, p_2, ..., p_s$ be theoretical probabilities, and $n_1, n_2, ..., n_s$ observed number of points fitting in corresponding intervals.

Thus, root form of chi-squared criterion is [5]:

$$4\left[n-\left(\sqrt{n_1 p_1}+\sqrt{n_2 p_2}+...+\sqrt{n_s p_s}\right)^2\right] = \chi^2_{s-1}. \tag{11}$$

Eq.11 means that if probability distribution corresponds with the theoretical one, then the left value is random of chi-squared form with $s-1$ degrees of freedom. Chi-squared standard form [1] follows from chi-squared criterion of the form (11) (as an asymptotic limit).

$s = 2$ case corresponds to binomial distribution, and root-form approximation by normal distribution is:

$$2\left(\sqrt{n_1 p_2} - \sqrt{n_2 p_1}\right) \sim N(0,1) \qquad (12)$$

where $p_1 + p_2 = 1$, $n_1 + n_2 = n$, $N(0,1)$ - is a random value of standard normal form.

Similar result of classical theory of probability is the Moivre- Laplace theorem (see [1]):

$$\frac{n_1 - np_1}{\sqrt{np_1 p_2}} \sim N(0,1) \qquad (13)$$

It is easy to ensure that eq. 13 asymptotically follows from eq.12. Nevertheless, for finite sample size approximation form (12) provides better accuracy compared to classical result (13). (see Fig.1, where the mean absolute error of root approximation is 1.82 times lower than the corresponding Moivre- Laplace error).

## 2. STATISTICAL ANALYSIS OF MUTUALLY COMPLEMENTING EXPERIMENTS

We have defined the psi function as a complex-valued function with the squared absolute value equal to the probability density. From this point of view, any psi function can be determined up to arbitrary phase factor $\exp(iS(x))$. In particular, the psi function can be chosen real-valued. At the same time, from the physical standpoint, the phase of psi function is not redundant. The psi function becomes essentially complex valued function in analysis of mutually complementing (according to Bohr) experiments with micro objects [6].

According to quantum mechanics, experimental study of statistical ensemble in coordinate space is incomplete and has to be completed by study of the same ensemble in another (canonically conjugate, namely, momentum) space. Note that measurements of ensemble parameters in canonically conjugate spaces (e.g., coordinate and momentum spaces) cannot be realized in the same experimental setup.

The uncertainty relation implies that the two-dimensional density in phase space $P(x, p)$ is physically senseless, since the coordinates and momenta of micro objects cannot be measured simultaneously. The coordinate $P(x)$ and momentum $\widetilde{P}(p)$ distributions should be studied separately in mutually complementing experiments and then combined by introducing the psi function.

The coordinate-space and momentum-space psi functions are related to each other by the Fourier transform

$$\psi(x) = \frac{1}{\sqrt{2\pi}} \int \widetilde{\psi}(p) \exp(ipx) dp \,, \quad \widetilde{\psi}(p) = \frac{1}{\sqrt{2\pi}} \int \psi(x) \exp(-ipx) dx \,. \qquad (14)$$

Consider a problem of estimating an unknown psi function ($\psi(x)$ or $\widetilde{\psi}(p)$) by experimental data observed both in coordinate and momentum spaces. We will refer to this problem as an statistical inverse problem of quantum mechanics [5,7-9] (do not confuse it with an inverse problem in the scattering theory). The predictions of quantum mechanics are considered as a direct problem. Thus, we consider quantum mechanics as a stochastic theory, i.e., a theory describing statistical (frequency) properties of experiments with random events. However, quantum mechanics is a special stochastic theory, since one has to perform mutually complementing experiments (space-time description has to be completed by momentum-energy one) to get statistically full description of a population (ensemble). In order for various representations to be mutually consistent, the theory should be expressed in terms of probability amplitude rather than probabilities themselves.

Methodologically, the method considered here essentially differs from other well known methods for estimating quantum states that arise from applying the methods of classical tomography and classical statistics to quantum problems [10-12]. The quantum analogue of the distribution density is the density matrix and the corresponding Wigner distribution function. Therefore, the methods developed so far have been aimed at reconstructing the aforementioned objects in analogy with the methods of classical tomography (this resulted in the term "quantum tomography") [13].

In [14], a quantum tomography technique on the basis of the Radon transformation of the Wigner function was proposed. The estimation of quantum states by the method of least squares was considered in [15]. The maximum likelihood technique was first presented in [16,17]. The version of the maximum likelihood method providing fulfillment of basic conditions imposed of the density matrix (hermicity, nonnegative definiteness, and trace of matrix equal to unity) was given in [18,19]. Characteristic features of all these methods are rapidly increasing calculation complexity with increasing number of parameters to be estimated and ill-posedness of the corresponding algorithms, not allowing one to find correct stable solutions.

The orientation toward reconstructing the density matrix overshadows the problem of estimating more fundamental object of quantum theory, i.e., the state vector (psi function). Formally, the states described by the psi function are particular cases of those described by the density matrix. On the other hand, this is the very special case that corresponds to fundamental laws in Nature and is related to the situation when the state described by a large number of unknown parameters may be stable and estimated up to the maximum possible accuracy.

Let us consider generalization of the maximum likelihood principle and likelihood equation for estimation of the state vector of a statistical ensemble on the basis of experimental data obtained in mutually complementing experiments. To be specific, we will assume that corresponding experiments relate to coordinate and momentum spaces.

We define the likelihood function as

$$L(x, p|c) = \prod_{i=1}^{n} P(x_i|c) \prod_{j=1}^{m} \widetilde{P}(p_j|c). \tag{15}$$

Here, $P(x_i|c)$ and $\widetilde{P}(p_j|c)$ are the densities in mutually complementing experiments corresponding to the same state vector $c$. We assume that $n$ measurements were made in the coordinate space; and $m$, in the momentum one.

Then, the log likelihood function has the form

$$\ln L = \sum_{i=1}^{n} \ln P(x_i|c) + \sum_{j=1}^{m} \ln \widetilde{P}(p_j|c). \tag{16}$$

The maximum likelihood principle together with the normalization condition evidently results in the problem of maximization of the following functional:

$$S = \ln L - \lambda(c_i c_i^* - 1), \tag{17}$$

where $\lambda$ is the Lagrange multiplier and

$$\ln L = \sum_{k=1}^{n} \ln(c_i c_j^* \varphi_i(x_k) \varphi_j^*(x_k)) + \sum_{l=1}^{m} \ln(c_i c_j^* \widetilde{\varphi}_i(p_l) \widetilde{\varphi}_j^*(p_l)). \tag{18}$$

Here, $\widetilde{\varphi}_i(p)$ is the Fourier transform of the function $\varphi_i(x)$.

Hereafter, we imply the summation over recurring indices numbering the terms of the expansion in terms of basis functions. On the contrary, statistical sums denoting the summation over the sample points will be written in an explicit form.

The necessary condition $\frac{\partial S}{\partial c_i^*} = 0$ for an extremum yields the likelihood equation

$$R_{ij} c_j = \lambda c_i \quad i,j = 0,1,\ldots,s-1, \tag{19}$$

where the $R$ matrix is determined by

$$R_{ij} = \sum_{k=1}^{n} \frac{\varphi_i^*(x_k)\varphi_j(x_k)}{P(x_k)} + \sum_{l=1}^{m} \frac{\widetilde{\varphi}_i^*(p_l)\widetilde{\varphi}_j(p_l)}{\widetilde{P}(p_l)}. \tag{20}$$

The problem (19) is formally linear. However, the matrix $R_{ij}$ depends on an unknown densities $P(x)$ and $\widetilde{P}(p)$. Therefore, the problem under consideration is actually nonlinear, and should be solved by the iteration method [5,7]. An exception is the histogram density estimator when the problem can be solved straightforwardly.

Multiplying both parts of Eq. (19) by $c_i^*$ and summing with respect to $i$, we find that the most likely state vector $c$ always corresponds to its eigenvalue $\lambda = n + m$ of the $R$ matrix (equal to sum of measurements).

An optimal number of harmonics in the expansion is appropriate to choose, on the basis of the compromise, between two opposite tendencies: the accuracy of the estimation of the function approximated by a finite series increases with increasing number of harmonics, however, the statistical noise level also increases.

The likelihood equation in the root state estimator method has a simple quasilinear structure and admits developing an effective fast-converging iteration procedure even in the case of multiparametric problems. The numerical implementation of the proposed algorithm is considered by the use of the set of Chebyshev-Hermite functions as a basis set of functions [5,7-8].

The implication of the root estimator method to statistical reconstruction of optical quantum states is considered in [9].

Examples of mutually complementing experiments that are of importance from the physical point of view are diffraction patterns (for electrons, photons, and any other particles) in the near-field zone (directly downstream of the diffraction aperture) and in the Fraunhofer zone (far from the diffraction aperture). The intensity distribution in the near-field zone corresponds to the coordinate probability distribution; and that in the Fraunhofer zone, the momentum distribution. The psi function estimated by these two distributions describes the wave field (amplitude and phase) directly at the diffraction aperture. The psi function dynamics described by the Schrödinger equation for particles and the Leontovich parabolic equation for light allows one to reconstruct the whole diffraction pattern (in particular, the Fresnel diffraction).

In the case of a particle subject to a given potential (e.g., an atomic electron) and moving in a finite region, the coordinate distribution is the distribution of the electron cloud, and the momentum distribution is detected in a thought experiment where the action of the potential abruptly stops and particles move freely to infinity.

In quantum computing, the measurement of the state of a quantum register corresponds to the measurement in coordinate space; and the measurement of the register state after performing the discrete Fourier transform, the measurement in momentum space. A quantum register involving $n$ qubits can be in $2^n$ states; and correspondingly, the same number of complex parameters is to be estimated. Thus, exponentially large number of measurements of identical registers is required to reconstruct the psi function if prior information about this function is lacking.

The state of quantum register is determined by the psi function

$$\psi = c_i |i\rangle \qquad (21)$$

The probability amplitudes in the conjugate space corresponding to complementing measurements are

$$\tilde{c}_i = U_{ij} c_j \qquad (22)$$

The likelihood function relating to $n+m$ mutually complementing measurements is

$$L = \prod_i (c_i c_i^*)^{n_i} \prod_j (\tilde{c}_j \tilde{c}_j^*)^{m_j} \qquad (23)$$

Here, $n_i$ and $m_j$ are the number of measurements made in corresponding states. In the case under consideration, the likelihood equation similar to (19) has the form

$$\frac{1}{n+m} \left[ \frac{n_i}{c_i^*} + \sum_j \frac{m_j U_{ji}^*}{\tilde{c}_j^*} \right] = c_i \qquad (24)$$

Figure 2 shows the comparison between exact densities that could be calculated if the psi function of an ensemble is known (solid line), and estimators obtained in mutually complementing experiments (quantum register: $8$ qubits, $2^8 = 256$ states). In each experiment, the sample size is 10000 points. In Fig. 3, the exact psi function is compared to that estimated by samples.

## 3. PROCESS AMPLITUDES AND EVENT GENERATION INTENSITY

The approach based on the use of psi-function is limited, in general by problems of non-relativity quantum mechanics. A more general approach is based on implementing a scattering matrix (S-operator)) [20]. Rigorously, problems of light interaction with matter, and photon field reconstruction, in particular, must be considered in this formalism framework [9]

Let $S$ - operator that sets transformation of in-state to out-state.

$$\Phi_{out} = S \Phi_{in} \qquad (25)$$

Let out-state be decomposed with a set of basis states

$$\Phi_{out} = c_j |j\rangle \qquad (26)$$

Experimental study of quantum out-state transits to the study of mutually complementary quantum processes. The processes' amplitude is

$$M_{ij} = \langle j|S|i\rangle \qquad (27)$$

Process amplitude square module specifies the intensity of event generation:

$$\lambda_j = M_j^* M_j \qquad (28)$$

The event-generation intensity $\lambda_j$ is the main quantity accessible for the measurement ($\lambda_j$ is measured in frequency units (Hz).). The number of events occurring in any given time interval obeys the Poisson distribution. Therefore, the quantities $\lambda_j$ specify the intensities of the corresponding mutually complementary Poisson processes and serve as estimates of the Poisson parameters (see below).

Although the amplitudes of the processes cannot be measured directly, they are of the greatest interest as quantities describing the fundamental relationships of quantum physics. From the superposition principle, it follows that the amplitudes are linearly related to the state-vector components. It is the purpose of quantum tomography to reproduce the amplitudes and state vectors which are hidden from the direct observation.

In some sense, the process amplitude is the "root" of the event generation intensity, likewise as ordinary psi-function is the "root" of probability density.

The linear transformation of the state vector $c$ into the amplitude of the process $M$ is described by a certain matrix $X$. Then the set of all amplitudes of the processes can be expressed by a single matrix equation

$$Xc = M \qquad (29)$$

We call the matrix $X$ the instrumental matrix of a set of mutually complementary measurements, by analogy with the conventional instrumental function. The matrix $X$ is known *a priori* (before the experiment). Concrete examples of instrumental matrices applied to the problems of quantum optics can be found in [9,21]

In eq.29 state vector is proposed to be non-normalized. The usage of non-normalized vector releases us from inserting an interaction constant in (29). The vector $c$ norm, obtained as the result of quantum system reconstruction, provides information of total intensity of all the processes considered in the experiment.

Now let us consider maximum likelihood estimator of state vector. The likelihood function is defined by the product of Poisson probabilities:

$$L = \prod_i \frac{(\lambda_i t_i)^{k_i}}{k_i!} e^{-\lambda_i t_i} \qquad (30)$$

where $k_i$ is the number of coincidences observed in the $i$ th process during the exposure time $t_i$, and $\lambda_i$ are the unknown theoretical event-generation intensities.

The log likelihood (logarithm of the likelihood function) is, except for an insignificant constant,

$$\ln L = \sum_i \left( k_i \ln(\lambda_i t_i) - \lambda_i t_i \right) \qquad (31)$$

We also introduce the matrices with elements defined by the following formulas:

$$I_{js} = \sum_i t_i X_{ij}^* X_{is} \qquad (32)$$

$$J_{js} = \sum_i \frac{k_i}{\lambda_i} X_{ij}^* X_{is} \quad i,s = 1,2,3 \qquad (33)$$

The matrix $I$ is determined from the experimental protocol. We shall call it hermitian matrix of Fisher information. On the contrary, the matrix $J$ is determined by the experimental values of $k_i$ and by the unknown event-generation intensities $\lambda_i$ Let us call it empirical matrix of Fisher information (see also Section 4).

In terms of these matrices, the condition for the extremum of function (20) can be written as

$$Ic = Jc \qquad (34)$$

whence it follows that

$$I^{-1}Jc = c \qquad (35)$$

We will call the latter relationship the likelihood equation. This is a nonlinear equation, because $\lambda_i$ depends on the unknown state vector $c$. Because of the simple quasi-linear structure, this equation can easily be solved by the iteration method [5,7]. The operator $I^{-1}J$ can be called quasi- identity operator. Note that it acts as the identical operator on only one vector in the Hilbert space, namely, on the vector corresponding to solution (35) and representing the maximum possible likelihood estimate for the state vector. The condition for existence of the matrix $I^{-1}$ is a condition imposed on the initial experimental protocol. The resulting set of equations automatically includes the normalization condition, which is written as

$$\sum_i k_i = \sum_i (\lambda_i t_i) \qquad (36)$$

This condition implies that, for all processes, the total number of detected events is equal to the sum of the products of event detection rates into the exposure time.

### 4. STATISTICAL FLUCTUATIONS OF STATE VECTOR OF QUANTUM SYSTEM

As already mentioned before, state vector with undefined primary norm provides the most complete information of the system. Fluctuations of quantum state (and norm fluctuations, in particular) in a normally functioning quantum information system should be within certain range, defined by statistical theory. This section is devoted to that problem.

Practical significance of accounting statistical fluctuations in quantum system deals with developing methods of estimation and control of precision and stability of quantum information system functioning, and also methods of detecting external interception to the system (Eve attack on the quantum channel between Alice and Bob).

The estimate of the non-normalized state vector $c$, obtained by the maximum likelihood principle, differs from the exact state vector $c^{(0)}$ by a random value $\delta c = c^{(0)} - c$. Let us consider statistical properties of the fluctuation vector $\delta c$ by expansion of the log likelihood function near the stationary point. The expansion is as follows:

$$-\delta \ln L = \left[ \frac{1}{2} \left( K_{sj} \delta c_s \delta c_j + K_{sj}^* \delta c_s^* \delta c_j^* \right) + I_{sj} \delta c_s^* \delta c_j \right], \qquad (37)$$

where together with the above (in (32)) defined hermitian matrix of Fisher information $I$, we define a symmetric Fisher information matrix $K$, which elements are defined by the following equation:

$$K_{sj} = \sum_v \frac{k_v}{M_v^2} X_{vs} X_{vj} \qquad (38)$$

where $M_\nu$ is the amplitude of the $\nu$ - th process.

$K$ in general case is a complex symmetric (non-hermitian) matrix.

From all the possible fluctuations let us mark out the so-called gauge fluctuations. Infinitesimal global gauge transformations of a state vector are as follows:

$$\delta c_j = i\varepsilon\, c_j, \quad j = 1,2,\ldots, s \qquad (39)$$

where $\varepsilon$ - is an arbitrary small real number, $s$ - the Hilbert space dimension.

Evidently, for gauge transformations $\delta \ln L = 0$. It means that two states vectors that differ by a gauge transformation, are statistically equivalent (have the same likelihood). Such vectors are equivalent physically too (global state vector phase is physically non-observable). From statistical point of view, the set of mutually complementing measurements should be chosen in a manner that for all the other fluctuations (except gauge) the equation (37) is strictly positive: $-\delta \ln L > 0$. We shall call this inequality the statistical completeness condition of a set mutually complementing measurement. Let us obtain constructive criteria of statistical completeness of measurements. The complex fluctuation vector $\delta c$ is convenient to be represented by a real vector of double length. Let us extract explicitly the real and the imaginary parts of the fluctuation vector $\delta c_j = \delta c_j^{(1)} + i\delta c_j^{(2)}$ and transit from the complex vector $\delta c$ to the real $\delta \xi$

$$\delta c = \begin{pmatrix} \delta c_1 \\ \delta c_2 \\ \delta c_3 \end{pmatrix} \rightarrow \delta \xi = \begin{pmatrix} \delta c_1^{(1)} \\ \vdots \\ \delta c_s^{(1)} \\ \delta c_1^{(2)} \\ \vdots \\ \delta c_s^{(2)} \end{pmatrix} \qquad (40)$$

In particular for qutrits ($s = 3$) this provides transition from 3-component complex vector to 6-component real vector.

In the new representation the equation (37) is expressed in the form:

$$\delta \ln L = -H_{sj}\delta\xi_s\delta\xi_j = -\langle \delta\xi | H | \delta\xi \rangle, \qquad (41)$$

where the matrix $H$ we shall call the complete information matrix. It is of the following block form:

$$H = \begin{pmatrix} \mathrm{Re}(I+K) & -\mathrm{Im}(I+K) \\ \mathrm{Im}(I-K) & \mathrm{Re}(I-K) \end{pmatrix} \qquad (42)$$

Matrix $H$ is real and symmetric. It is of double dimension to matrices $I$ and $K$.

Now for one it is easy to formulate the desired characteristic condition of mutually complementing measurement set completeness. For a set of measurements to be statistically complete, it is necessary and sufficient that one (and the only one) eigenvalue of the complete information matrix H is equal to zero, while the other are strictly positive.

Notice that by checking the condition, one not only verifies statistical completeness of a measurement protocol but also, insures that the obtained extremum is of maximum likelihood.

Eigenvector that has eigenvalue equal to zero corresponds to gauge fluctuation direction (such fluctuations are of no physical importance as stated above). Eigenvectors corresponding to the other eigenvalues specify in Hilbert space directions that we shall call principle state vector fluctuation directions.

Principle fluctuations variance is

$$\sigma_j^2 = \frac{1}{2h_j}, \qquad j = 1,\ldots,2s-1 \qquad (43)$$

where $h_j$ is the eigenvalue of the information matrix $H$, corresponding to the $j$-the principle direction.

The most critical direction in Hilbert space is the one with the maximum variance $\sigma_j^2$, while the eigenvalue $h_j$ is accordingly minimal. The knowledge of numeric dependences of statistical fluctuations allows researcher to estimate distributions of various statistical characteristics.

The most important information criterion that specifies the general possible level of statistical fluctuations in quantum information system is the chi-square criterion. According to the stated above it can be expressed as:

$$2\langle \delta\xi | H | \delta\xi \rangle \sim \chi^2(2s-1) \qquad (44)$$

where $s$ is the Hilbert space dimension

Equation (44) has the meaning that the left value, that describes the level of state vector information fluctuations is of chi-square distribution with $2s-1$ degrees of freedom.

The validity of the analytical expression (44) is justified by the results of numerical modeling and observed data (see [21]). Similarly to (40) let us introduce the transformation of a complex state vector to a real vector of double length:

$$c = \begin{pmatrix} c_1 \\ c_2 \\ c_3 \end{pmatrix} \rightarrow \xi = \begin{pmatrix} c_1^{(1)} \\ \vdots \\ c_s^{(1)} \\ c_1^{(2)} \\ \vdots \\ c_s^{(2)} \end{pmatrix} \qquad (45)$$

It can be shown that the information carried by a state vector is equal to double total number of observations in all processes.

$$\langle \xi | H | \xi \rangle = 2n, \qquad (46)$$

where $n = \sum_v k_v$

Then, the chi-square criterion can be expressed in the form invariant to the state vector scale (let us remind that we consider a non-normalized state vector).

$$\frac{\langle \delta\xi | H | \delta\xi \rangle}{\langle \xi | H | \xi \rangle} \sim \frac{\chi^2(2s-1)}{4n} \qquad (47)$$

Equation (47) describes distribution of relative informational fluctuation. It shows that relative information uncertainty of a quantum state decreases with number of observations as $1/n$. The mean value of relative information fluctuation is:

$$\overline{\frac{\langle \delta\xi | H | \delta\xi \rangle}{\langle \xi | H | \xi \rangle}} = \frac{2s-1}{4n} \qquad (48)$$

As a measure of correspondence of a theoretical state vector and its' estimate let us introduce a characteristic, that we shall call informational fidelity.

$$F_H = 1 - \frac{\langle \delta\xi | H | \delta\xi \rangle}{\langle \xi | H | \xi \rangle} \qquad (49)$$

Value $1 - F_H$ we shall call informational loss.

The convenience of informational fidelity $F_H$ is enclosed in its' simpler statistical properties compared to the conventional one $F$. For a system where statistical fluctuations are dominant fidelity is a random value, based on chi-square distribution.

$$F_H = 1 - \frac{\chi^2(2s-1)}{4n}, \qquad (50)$$

where $\chi^2(2s-1)$ is a random value of chi-square type with $2s-1$ degrees of freedom.

Informational fidelity value asymptotically tends to unity with sample size growth, while informational loss tends to zero. Complementary (to statistical fluctuations) noise decreases informational fidelity level compared to the theoretical level (50).

The examples of applying the theory to quantum optical state reconstruction can be found in [9,21].

### 5. ROOT ESTIMATOR AND QUANTUM DYNAMICS

Assume that the mechanical equations are satisfied only for statistically averaged quantities (the averaged Newton's second law of motion)

$$\frac{d^2}{dt^2}\left(\int P(x)\vec{x}\,dx\right) = -\frac{1}{m}\left(\int P(x)\frac{\partial U}{\partial \vec{x}}\,dx\right) \qquad (51)$$

Let us require the density $P(x)$ to admit the root expansion [8] ($s$ components), i.e.,

$$P(x) = |\psi^{(1)}(x)|^2 + |\psi^{(2)}(x)|^2 + \ldots + |\psi^{(s)}(x)|^2, \qquad (52)$$

where

$$\psi^{(l)}(x) = c_j^{(l)}(t)\varphi_j(x) \qquad l = 1,\ldots,s \qquad (53)$$

We will search for the time dependence of the expansion coefficients in the form of harmonic dependence

$$c_j^{(l)}(t) = c_{j0}^{(l)} \exp(-i\omega_j t). \qquad (54)$$

Then, Eq. (51) yields

$$m(\omega_j - \omega_k)^2 \sum_{l=1}^{s} c_{j0}^{(l)} c_{k0}^{*(l)} \langle k|\vec{x}|j\rangle \exp(-i(\omega_j - \omega_k)t) =$$
$$= \sum_{l=1}^{s} c_{j0}^{(l)} c_{k0}^{*(l)} \langle k|\frac{\partial U}{\partial \vec{x}}|j\rangle \exp(-i(\omega_j - \omega_k)t) \qquad (55)$$

Here, the summation over recurring indices $j$ and $k$ is meant. The matrix elements in (55) are determined by the formulas

$$\langle k|\vec{x}|j\rangle = \int \varphi_k^*(x)\,\vec{x}\,\varphi_j(x)\,dx \qquad (56)$$

$$\langle k|\frac{\partial U}{\partial \vec{x}}|j\rangle = \int \varphi_k^*(x)\frac{\partial U}{\partial \vec{x}}\varphi_j(x)dx \quad (57)$$

In order for the expression (55) to be satisfied at any instant of time for arbitrary initial amplitudes, the left and right sides are necessary to be equal for each matrix element. Therefore,

$$m(\omega_j - \omega_k)^2 \langle k|\vec{x}|j\rangle = \langle k|\frac{\partial U}{\partial \vec{x}}|j\rangle \quad (58)$$

This expression is a matrix equation of the Heisenberg quantum dynamics in the energy representation (written in the form similar to that of the Newton's second law of motion). The basis functions and frequencies satisfying (58) are the stationary states and frequencies of a quantum system, respectively (in accordance with the equivalence of the Heisenberg and Schrödinger pictures).

Indeed, let us construct the diagonal matrix from the system frequencies $\omega_j$. The matrix under consideration is Hermitian, since the frequencies are real numbers. This matrix is the representation of a Hermitian operator with eigenvalues $\omega_j$, i.e.,

$$H|j\rangle = \hbar\omega_j|j\rangle \quad (59)$$

Let us find an explicit form of this operator. In view of (59), the matrix relationship (58) can be represented in the form of the operator equation

$$[H[Hx]] = \frac{\hbar^2}{m}\hat{\partial}U, \quad (60)$$

where $\hat{\partial} = \frac{\partial}{\partial x}$ is the operator of differentiation and $[\ ]$, the commutator.

The Hamiltonian of a system

$$H = -\frac{\hbar^2}{2m}\hat{\partial}^2 + U(x) \quad (61)$$

is the solution of operator equation (60).

Let us consider density matrix with the elements:

$$\rho_{jk} = \sum_{l=1}^{s} c_j^{(l)} c_k^{*(l)} = \sum_{l=1}^{s} c_{j0}^{(l)} c_{k0}^{*(l)} \exp(-i(\omega_j - \omega_k)t) \quad (62)$$

Basing on the above results one can easily derive the equation for density matrix dynamics, usually called quantum Liouville equation.

$$\frac{\partial \rho}{\partial t} = -\frac{i}{\hbar}[H, \rho] \quad (63)$$

Thus, if the root density estimator is required to satisfy the averaged classical equations of motion, the basis functions and frequencies of the root expansion cannot be arbitrary, but have to be eigenfunctions and eigenvalues of the system Hamiltonian, respectively.

The relationships providing that the averaged equations of classical mechanics are satisfied for quantum systems are referred to as the Ehrenfest equations [22]. These equations are insufficient to describe quantum dynamics. As it has been shown above, an additional condition allowing one to transform a classical system into the quantum one (i.e., quantization condition) is actually the requirement for the density to be of the root form.

Thus, if we wish to turn from the rigidly deterministic (Newtonian) description of a dynamical system to the statistical one, it is natural to use the root expansion of the density distribution to be found, since only in this case a stable statistical model can be found. On the other hand, the choice of the root expansion basis determined by the eigenfunctions of the energy operator (Hamiltonian) is not simply natural, but the only possible way consistent with the dynamical laws.

## CONCLUSIONS

Let us state a short summary.

Search for multiparametric statistical model providing stable estimation of parameters on the basis of observed data results in constructing the root density estimator. The root density estimator is based on the representation of the probability density as a squared absolute value of a certain function, which is referred to as a psi-function in analogy with quantum mechanics. The method proposed is an efficient tool to solve the basic problem of statistical data analysis, i.e., estimation of distribution density on the basis of experimental data.

The coefficients of the psi-function expansion in terms of orthonormal set of functions are estimated by the maximum likelihood method providing optimal asymptotic properties of the method (asymptotic unbiasedness, consistency, and asymptotic efficiency). The introduction of the psi-function allows one to represent the Fisher information matrix as well as statistical properties of the sate vector estimator in simple analytical forms. Basic objects of the theory (state vectors, information and covariance matrices etc.) become simple geometrical objects in the Hilbert space that are invariant with respect to unitary (orthogonal) transformations.

A new statistical characteristic, a confidence cone, is introduced instead of a standard confidence interval. The chi-square test is considered to test the hypotheses that the estimated vector equals to the state vector of general population.

The root state estimator may be applied to analyze the results of experiments with micro objects as a natural instrument to solve the inverse problem of quantum mechanics: estimation of state vector by the results of mutually complementing (according to Bohr) measurements (processes). Generalization of the maximum likelihood principle to the case of statistical analysis of mutually complementing experiments is proposed.

It is shown that the requirement for the density to be of the root form is the quantization condition. Actually, one may say about the root principle in statistical description of dynamic systems. According to this principle, one has to perform the root expansion of the distribution density in order to provide the stability of statistical description. On the other hand, the root expansion is consistent with the averaged laws of classical mechanics when the eigenfunctions of the energy operator (Hamiltonian) are used as basis functions. Figuratively speaking, there is no a regular statistical method besides the root one, and there is no regular statistical mechanics besides the quantum one.

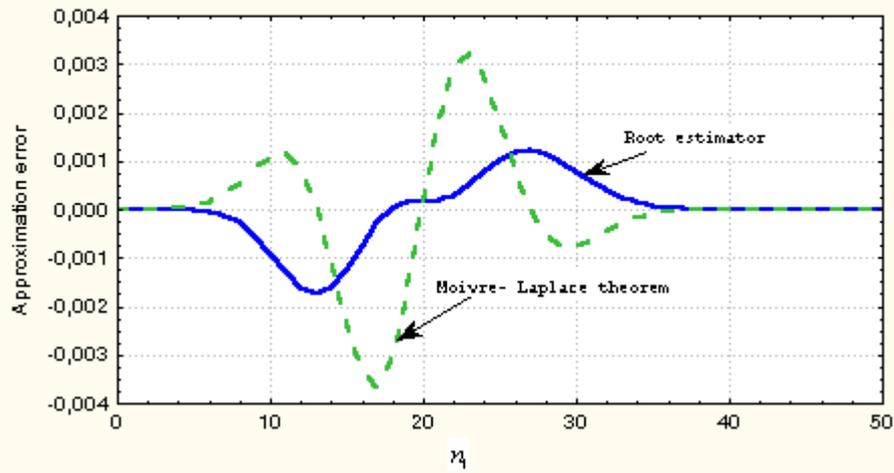

Fig.1 A compare of root estimator and Moivre-Laplace theorem when approximating binomial distribution by normal

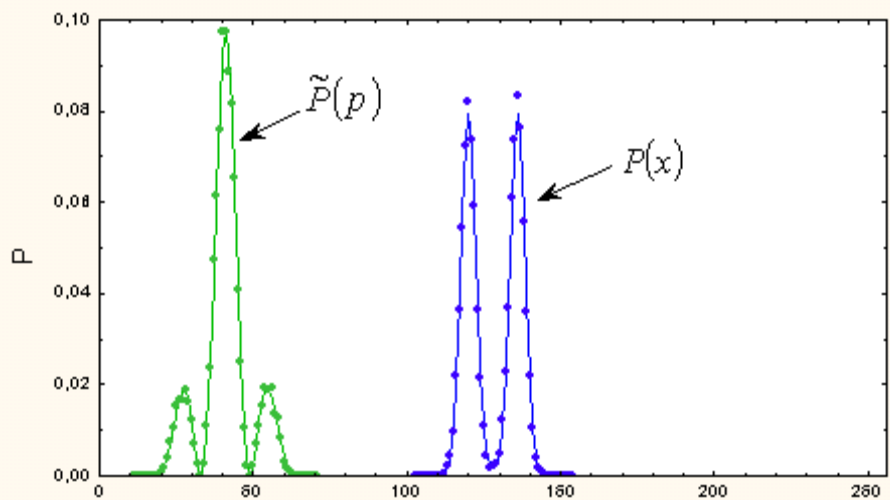

Fig.2 Mutually complementing distributions (coordinate and momentum probability distributions)

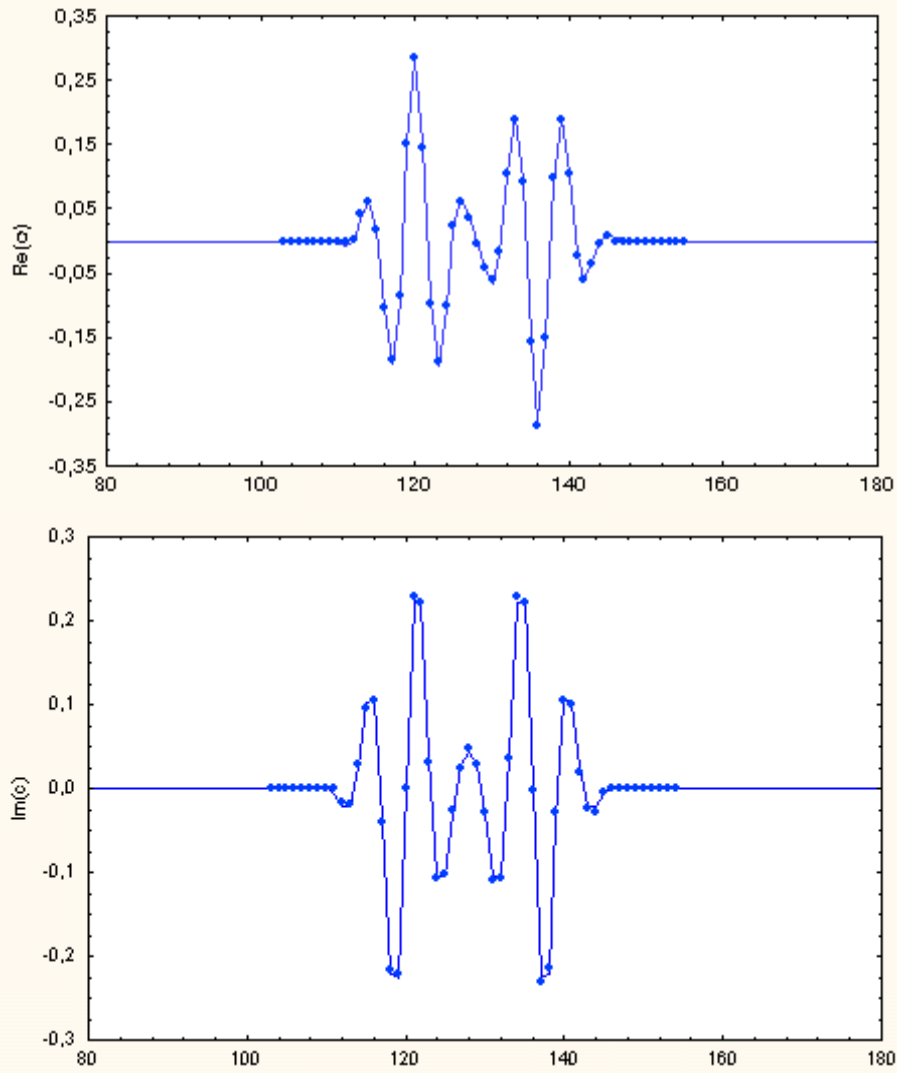

Fig. 2 Comparison between exact psi-function (solid line) and that estimated by a sample (dots)